\documentstyle[11pt]{article}
\begin{document}
\title{The entanglement criterion of multiqubits}
\author{Hyuk-jae Lee{\footnote{e-mail:lhjae@iquips.uos.ac.kr}}\\
Institute of Quantum Information Processing and Systems,\\
University of Seoul, Seoul, 130-743, Korea\\
\and Sung Dahm Oh$^2${\footnote{e-mail:sdoh@sookmyung.ac.kr}}\\
Department of Physics,\\
Sookmyung Women's University, Seoul,140-742, Korea\\
An associate member of Korea Institute for Advanced study, Seoul,
Korea\\
\and Doyeol Ahn{\footnote{e-mail:dahn@uoscc.uos.ac.kr}}\\
Institute of Quantum Information Processing and Systems,\\
University of Seoul, Seoul, 130-743, Korea,\\
Department of Electrical and Computer Engineering,\\
University of Seoul, Seoul, 130-743, Korea}

%
\maketitle

\begin{abstract}
We present an entanglement criterion for
multiqubits by using the quantum correlation tensors which rely on the expectation values of the Pauli operators
for a multiqubit state. Our criterion explains not only the total
entanglement of the system but also the partial entanglement
in subsystems. It shows that we have to consider the
subsystem entanglements in order to obtain the full description
for multiqubit entanglements. Furthermore, we offer an extension
of the entanglement to multiqudits.
\end{abstract}
\vspace{.25in}


\newpage
Entanglement has been an important key word in the quantum computer and the quantum information
technology. In particular, entanglements in the bipartite qubits have many applications as
the superdense coding\cite{benn}, quantum computation, teleportation\cite{bras}, clock
synchronization\cite{chua},\cite{abra}and quantum cryptography\cite{eker}. These
have been clarified by a negative partial transposition \cite{pere},\cite{horo}, and
quantified by concurrence\cite{woot},
negativity\cite{vida}, entanglement of formation\cite{benne}, etc..

As many entangled states such as GHZ state, W state, etc., were found in multiqubits, the multiqubit
entanglements have been applied to a real physical system such as quantum secret sharing and the one-way quantum computer\cite{brie}.
The investigation on the entanglement properties in the multipartite system has thus emerged as a central
problem in quantum information study. However, no efficient method to clarify the status of multipartite entanglements has been introduced.

The classification of the mathematical and physical
structures in multipartite states was, at first, tried by the
local operation associated with classical communication(LOCC).  In the
multipartite systems, the investigations on entanglement measure
have been proposed by the {\it tangle}\cite{coff} which is
computed by the concurrence between two intentionally divided
subsystems in an effective two-dimensional Hilbert space.
Recently, entanglement witnesses were suggested by another method
for the classification of multipartite entanglements\cite{horo},\cite{bour}. This method requires the
witness operators to detect various forms of multipartite entanglements. However, it is difficult to know the witness
operators before we classify the multipartite system, and additionally witness operators are defined by some a priori
knowledge about the states under investigation.

In spite of trials, the entanglements in a multipartite system are complicated even in pure systems because
the quantum states can share entanglements differently among
possible subsystems and have the different classes of total entangled states as GHZ, W or cluster states.
It is important to define an entanglement criterion
that could distinguish all possible types of
entanglements which exist among the constituents. In this letter,
we present a general entanglement criterion that can solve the
above problems for pure multiqubit systems. Furthermore we will
discuss that our entanglement description can be extended to
multiqubit mixed states and higher dimensional Hilbert spaces.

In general, a pure composite system with $N$ qubits can be represented
by
\begin{equation}
|\Psi(1, 2, 3,\cdots,N)\rangle=\sum^1_{IJ\cdots K=0}a_{IJ\cdots K}
|I\rangle_1 \otimes|J\rangle_2\otimes\cdots\otimes|K\rangle_N,
\label{abc}
\end{equation}
where $\sum^1_{IJ\cdots K=0}|a_{IJ\cdots K}|^2 =1$ and the state is an element of composite Hilbert space as ${\cal H}_N={\cal H}_2\otimes
{\cal H}_2\otimes\cdots \otimes{\cal H}_2$.

Our question is whether this state is separable or entangled. In
order to answer this question, we introduce the quantum
correlation tensor\cite{mahl} for the given multipartite qubit
system $|\Psi\rangle$ as
\begin{eqnarray}
&&M_{i_1 i_2\cdots i_n}(\alpha_1, \alpha_2, \cdots, \alpha_N
;|\Psi\rangle)=
\langle\Psi|(\sigma_{i_1}(\alpha_1)-\lambda_i(\alpha_1))\otimes(\sigma_{i_2}(\alpha_2)-\lambda_{i_2}(\alpha_2))\nonumber\\
&&\otimes\cdots\otimes(\sigma_{i_n}(\alpha_n)-\lambda_{i_n}(\alpha_n))\otimes
I(\alpha_{n+1})\cdots \otimes I(\alpha_N)|\Psi\rangle,
\label{smtensor}
\end{eqnarray}
where $n\le N$, $\sigma_i(\alpha)$ denotes the $i^{th}$-component
Pauli's operator of the $\alpha^{th}$ qubit and
$\lambda_{i_j}(\alpha_j)=\langle\Psi| I(1)\otimes I(2)\otimes
\cdots \sigma_{i_j}(\alpha_j)\cdots \otimes I(N)|\Psi\rangle$.
Here, $I(\alpha)$ is the identity operator on the $\alpha^{th}$
qubit. Obviously, $M$ determines whether one qubit is
separated. If the state is separable such as $|\Psi(1, 2,
3,\cdots,N)\rangle= |\Psi(\alpha_1, \alpha_2, \cdots,
\alpha_{N-1})\rangle \otimes|\Psi(\alpha_N)\rangle$, $M_{i_1
i_2\cdots i_N}(1, 2, \cdots, N;|\Psi\rangle)$ must be zero. This
fact can be proved by a simple calculation. Conversely, if $M_{i_1
i_2\cdots i_N}(1, 2, \cdots, N;|\Psi\rangle)=0$ for all $i_1, i_2,
\cdots, i_N$, the state is separable such as $|\Psi(1, 2,
3,\cdots,N)\rangle= |\Psi(\alpha_1, \alpha_2, \cdots,
\alpha_{N-1})\rangle \otimes|\Psi(\alpha_N)\rangle$, which implies
that one of the qubits is uncorrelated with the others.
We can consider a $N$-qubit system as a bipartite system consisted
of a $(N-1)$-qubit system and a single qubit system. So the
original state can be rewritten by the Schmidt's decomposition as
\begin{equation}
|\Psi(1,2,3,\cdots,N)\rangle=\alpha|a, 0\rangle+\beta|b, 1\rangle,
\end{equation}
where $|a\rangle$ and $|b\rangle$ are orthogonal states of the
$(N-1)$-qubit system with $\alpha^2+\beta^2=1$. We denote the
operators of the $(N-1)$-qubit space as a generator, $\hat{L_K}$.
We can calculate $M$ for a bipartite system as following;
\begin{eqnarray}
M_{Kx}&=&(\alpha\langle a,0|+\beta\langle b,1|)\hat{L_K}\otimes \sigma_x(\alpha|a, 0\rangle+\beta|b, 1\rangle)-L_K \lambda_x\nonumber\\
&=&2\alpha\beta Re(\langle a|\hat{L_K}|b\rangle),\nonumber\\
M_{Ky}&=&(\alpha\langle a,0|+\beta\langle b,1|)\hat{L_K}\otimes \sigma_y(\alpha|a, 0\rangle+\beta|b, 1\rangle)-L_K \lambda_y\nonumber\\
&=&2\alpha\beta Im(\langle a|\hat{L_K}|b\rangle),\nonumber\\
M_{Kz}&=&(\alpha\langle a,0|+\beta\langle b,1|)\hat{L_K}\otimes \sigma_z(\alpha|a, 0\rangle+\beta|b, 1\rangle)-L_K \lambda_z\nonumber\\
&=&\alpha^2\langle a|\hat{L_K}|a\rangle-\beta^2\langle b|\hat{L_K}|b\rangle
-(\alpha^2-\beta^2)(\alpha^2\langle a|\hat{L_K}|a\rangle+\beta^2\langle b|\hat{L_K}|b\rangle),\nonumber
\end{eqnarray}
where $L_K=(\alpha\langle a,0|+\beta\langle b,1|)(\hat{L_K}\otimes
I)(\alpha|a, 0\rangle+\beta|b, 1\rangle)$. $\alpha$ or $\beta$ has
to be vanished in order for all of $M$'s to be zero.
This indicates that at least one of the qubits is uncorrelated with
the others.

For a two-qubit system $M_{ij}$ is the criterion to judge whether
the bipartite pure state is separated or entangled. Schlinz and
Mahler suggested this scenario in a bipartite system\cite{mahl}.
The three-qubit state, $|\Psi(1,2,3)\rangle$, has three types of
entanglements; a separated state as $A-B-C$, a bipartite
entangled state as $A-BC, AB-C$ or $C-AB$ and a totally entangled state as
$ABC$. Nonzero of any
$M_{ijk}(1,2,3;|\Psi\rangle)$ tells us that the state is totally
entangled as the GHZ or W state. Zero of
$M_{ijk}(1,2,3;|\Psi\rangle)$ for any $i, j, k$ means that the
state can be either $|\Psi(1,2)\rangle\otimes
|\Psi(3)\rangle$ or $|\Psi(1)\rangle\otimes |\Psi(2)\rangle\otimes
|\Psi(3)\rangle$. However, the increase of qubit numbers in the system
produces the increase of the possibilities in the entanglement
types. We have to differentiate all these situations in order to
fully describe the entanglement structure. When $M_{i_1 i_2
i_3\cdots i_N}(\alpha_1, \alpha_2, \cdots, \alpha_N
;|\Psi\rangle)=0$ for all $i_1, i_2, \cdots, i_N$, there are
various situations including completely separable and
multi-separable cases. We can easily investigate that the state is
at least one-qubit separated such as $|\Psi(\alpha_1, \alpha_2,
\cdots, \alpha_{N-1})\rangle \otimes|\Psi(\alpha_N)\rangle$, but
we cannot judge directly whether the state, $|\Psi(\alpha_1,
\alpha_2, \cdots, \alpha_{N-1})\rangle$ is entangled or separated.
$M_{i_1 i_2\cdots i_n}(\alpha_1, \alpha_2, \cdots, \alpha_n
;|\Psi\rangle)$ with $n<N$ needs to determine the entanglement of
the subsystems consisted of $n$ qubits. $M_{i_1 i_2\cdots
i_n}(\alpha_1, \alpha_2, \cdots, \alpha_n;|\Psi\rangle)$ is not
zero for the states which have the entanglement among the $n$ qubits.
$M_{i_1 i_2\cdots i_n}(\alpha_1, \alpha_2, \cdots, \alpha_n
;|\Psi\rangle)$ can distinguish totally entangled state from
partially entangled state such as $|\Psi(\alpha_{i_1},
\alpha_{i_2}, \cdots, \alpha_{i_n})\rangle
\otimes|\Psi(\alpha_{i+1})\rangle\otimes|\Psi(\alpha_{i+2})\rangle\otimes\cdots\otimes|\Psi(\alpha_{N})\rangle$.

The correlation tensor, $M$, is sufficient in bipartite and
tripartite systems. However, the situation is different in systems
consisted of more than three qubits. They cannot assure a
criterion for partially entangled cases such as $|\Psi(\alpha_1,
\alpha_2, \cdots, \alpha_l)\rangle
\otimes|\Psi(\beta_1,\beta_2 \cdots \beta_m)\rangle\otimes\cdots\otimes|\Psi(\gamma_1,\gamma_2 \cdots \gamma_n)$.
We have to modify the quantum correlation tensors to solve the
problem by
\begin{eqnarray}
&&M'_{i_1 i_2\cdots i_n}(\alpha_1, \alpha_2, \cdots, \alpha_n;|\Psi\rangle)
= M_{i_1 i_2\cdots i_n}(\alpha_1, \alpha_2, \cdots, \alpha_n;|\Psi\rangle)\nonumber\\
&&-\sum_{A\cup B\cup\cdots\cup C=\{1,2,3\cdots n\}} (M_A( \alpha'_1, \alpha'_2, \cdots, \alpha'_{n_1};|\Psi\rangle)
M_B(\alpha''_1, \alpha''_2, \cdots, \alpha''_{n_2};|\Psi\rangle)\nonumber\\
&&\cdots M_C(\alpha'''_1, \alpha'''_2, \cdots, \alpha'''_{n_m};|\Psi\rangle),
\label{mtensorp}
\end{eqnarray}
where $n_1 +n_2 +\cdots+ n_m=n$ and $A=i_1 i_2\cdots i_{n_1}$ and
$i_j$ denotes the $i^{th}$-component of Pauli' operator acted on
the $j^{th}$ qubit. The sum of the second term in the right side of
eq. (\ref{mtensorp}) denotes the possible disjoint partitions of
indices composed of Pauli's components of each qubit. Since $M'$
equals to $M$ in bipartite and tripartite systems, it is
requested in the systems which consist of more than three qubits.
For instance, $M'$ of the four-qubit case is written by
\begin{eqnarray}
&&M'_{i_1 i_2 i_3 i_4}(1,2,3,4;|\Psi\rangle)=M_{i_1 i_2 i_3 i_4}(1,2,3,4;|\Psi\rangle)-M_{i_1 i_2}(1,2;|\Psi\rangle)M_{i_3 i_4}(3,4;|\Psi\rangle)\nonumber\\
&&-M_{i_1 i_3}(1,3;|\Psi\rangle)M_{i_2 i_4}(2,4;|\Psi\rangle)-M_{i_1 i_4}(1,4;|\Psi\rangle)M_{i_2 i_3}(2,3;|\Psi\rangle).
\label{qii}
\end{eqnarray}

This provides the complete criterion for multipartite entanglement
which includes the multi-separable subsystems. If $M'_{i_1
i_2\cdots i_n}(\alpha_1, \alpha_2, \cdots,
\alpha_N;|\Psi\rangle)=0$ for any $i_1, i_2,\cdots, i_n$, the given
state is separable, and has two possibilities. The first is that every term on the right hand side in eq. (\ref{qii}) is zero.
This means that the state has the from of $|\Psi(1,2,3,4)\rangle=|\psi(\alpha_1, \alpha_2, \alpha_3)\rangle\otimes|\phi(\alpha_4)\rangle$.
The second is that the first term, $M_{i_1 i_2 i_3 i_4}(1,2,3,4;|\Psi\rangle)$, is
subtracted by three terms with a negative sign. However, we can show without difficulties that
only one term out of three terms with the negative sign is nonzero; if
$M_{ij}(1,2,;|\Psi\rangle)M_{kl}(3,4;|\Psi\rangle)\ne 0$ for the
four-qubit case, $M_{i_1 i_2 i_3 i_4}(1,2,3,4;|\Psi\rangle)=M_{ij}(1,2;|\Psi\rangle)M_{kl}(3,4;|\Psi\rangle)$.
This fact lead us to find that the state is $|\Psi(1, 2, 3, 4)\rangle=|\psi(1,
2)\rangle\otimes |\phi(3, 4)\rangle$.

If any pure state of a $N$-qubit system is given, we must, at
first, check whether $M'_{i_1 i_2\cdots i_N}(\alpha_1, \alpha_2,
\cdots, \alpha_N;|\Psi\rangle)$ is zero or not. In the nonzero
case, the $N$-qubit is totally entangled but in the zero case,
the state is either completely separated or partially entangled, as described earlier.
In the zero case, we have to make additional checks whether
$M'_{i_1, i_2, \cdots \i_n}(\alpha_1, \alpha_2, \cdots, \alpha_n;|\Psi\rangle)$ for $n<N$ is zero or not in sequence.
These sequential checks determine whether the given state is totally entangled, biseparable, triseparable, $\cdots$ or completely
separable. Then $M'$ classifies all the possible forms of entangled states.

By the tensor, $M'$, we can classify the pure multiqubit states
including many different entanglement types. However,
the tensor cannot distinguish the entangled states which are connected to each others
under a local unitary transformation. For instance,
$M'$ can distinguish the product state from the Bell states in a
two-qubit system. However, the values of $M'$ of four Bell states are
different from each others. One may misunderstand that the four
Bell states have different degrees of entanglements. It is well known that
the four Bell states are equivalent under a local
unitary transformation as maximally entangled states.
$M'$ just distinguishes whether the multiqubit state are entangled or
multiseparated. Therefore, we need a new quantity to determine an entanglement magnitude.

Based on $M'$, we introduce a measure of an entanglement such as
\begin{equation}
B^{(m)}(\alpha_1, \alpha_2,\cdots
\alpha_m;|\Psi\rangle)=\sum_{ijkl\cdots}M'_{ijkl\cdots}(\alpha_1,
\alpha_2,\cdots \alpha_m;|\Psi\rangle)M'_{ijkl\cdots}(\alpha_1,
\alpha_2,\cdots \alpha_m;|\Psi\rangle). \label{entmeasure}
\end{equation}
$B^{(m)}(\alpha_1, \alpha_2, \cdots\alpha_m;|\Psi\rangle)$
calculates the entanglement magnitude among $m$ qubits labeled by
$\alpha_1, \alpha_2, \cdots\alpha_m$. For example,
$B^{(2)}(\alpha_1, \alpha_2;|\Psi\rangle)$ describes the
entanglement magnitude between the qubits $\alpha_1$ and $\alpha_2$
and $B^{(3)}(\alpha_1, \alpha_2, \alpha_3;|\Psi\rangle)$ the
entanglement degree among the qubits $\alpha_1, \alpha_2, \alpha_3$.
$B^{(2)}(1,2;|\Psi\rangle)$ in two-qubit systems is the same
measure as Schlinz and Mahler's entanglement measure\cite{mahl}.

$B^{(m)}$ of eq. (\ref{entmeasure})  satisfies the following
properties affirming entanglement monotone,
First, it is nonnegative, because $B$ is defined by the square of real numbers and
$M'=0$ in separable states. Second, it is invariant under any local unitary transformations. This can be shown easily
by using $U^{\dagger}\sigma_i
U=T_{ij}\sigma_j$ and $\sum_{i}T_{ij}T_{ik}=\delta_{jk}$ where $U$
is a unitary matrix and $T$ is a $3\times 3$ orthogonal matrix in
the qubit systems\cite{mahl}. Third, it is nonincreasing under local measurements.
The measurement collapses an entangled n-qubit state to
biseparable state that one qubit under local measurement processes
are uncorrelated with the others. Then, $B^{(m)}$ is vanished
under local measurements and less than that of the original state.
Fourth, it is invariant under the addition of an uncorrelated ancillary state. Fifth,
it is not increased by tracing out a part of the system.
Thus we claim that $B^{(m)}$ is the entanglement monotone.

For example, we consider four-qubit states which are totally entangled,
\begin{eqnarray}
|GHZ_4\rangle &=&\frac{1}{{\sqrt 2}}(|0000\rangle+|1111\rangle,\nonumber\\
|W_4\rangle &=&\frac{1}{2}(|0001\rangle+|0010\rangle+|0100\rangle+|1000\rangle),\nonumber\\
|\phi_6\rangle &=&\frac{1}{\sqrt{6}}(|0011\rangle+|0101\rangle+|1001\rangle+|1010\rangle+|0110\rangle+|1100\rangle)\nonumber\\
|\phi_4\rangle &=&\frac{1}{2}(|0000\rangle+|0011\rangle+|1100\rangle-|1111\rangle).
\end{eqnarray}
We can get $M'_{ijkl}(1,2,3,4;|\Psi\rangle)\ne 0$ for the above four states and know that their states are totally entangled.
Then we can distinguish the entanglement difference through calculation of $B^{(4)}$;
$B^{(4)}(1,2,3,4;|GHZ_4\rangle)=1, B^{(4)}(1,2,3,4;|W_4\rangle)=\frac{51}{256}, B^{(4)}(1,2,3,4;|\phi_6\rangle)=\frac{7}{27}$, and
$B^{(4)}(1,2,3,4;|\phi_4\rangle)=\frac{1}{3}$. Here we normalized $B^{(4)}$ with the value of the GHZ state.
A partially three-qubit entangled state as $|GHZ_3\rangle\otimes |0\rangle$ and a partially two-qubit entangled state as
$|Bell_2\rangle\otimes|Bell_2\rangle$ where $|Bell_2\rangle=\frac{1}{2}(|00\rangle+|11\rangle)$ have $M'=0$ for all subindexes.
Then this explains these states do not have total entanglement.
However, we can obtain $M'(1, 2, 3)$ is not zero for $|GHZ_3\rangle\otimes |0\rangle$ and  $M'(1,2)$ and $M'(3,4)$ are not zero for
$|Bell_2\rangle\otimes|Bell_2\rangle$. This shows that $|GHZ_3\rangle\otimes |0\rangle$ has a partial entanglement
among qubits $1, 2$ and $3$ and $|Bell_2\rangle\otimes|Bell_2\rangle$ has partial entanglements between qubits
$1$ and $2$, and between qubits $3$ and $4$.

If the Pauli's operators of the eq. (\ref{smtensor}) which are the
generators of $SU(2)$ unitary group are replaced by the generators
related to the higher dimensional Hilbert space, the same method
can be also applied to the higher dimensional composite systems.

Here we have mainly focused on the pure systems.
In the mixed case, it is not easy to apply directly as
\begin{eqnarray}
M_{i_1 i_2\cdots i_n}(1, 2, \cdots, n ;\rho)&=& tr[ \rho
(\sigma_{i_1}(1)-\lambda_i(1))\otimes(\sigma_{i_2}(2)-\lambda_{i_2}(2))\otimes\cdots\nonumber\\
&&\otimes(\sigma_{i_n}(n)-\lambda_{i_n}(n))], \label{m1tensor}
\end{eqnarray}
where $\rho$ is the density operator. In the Werner state of
two-qubit parameterized by fidelity with the singlet states, we
get $M'\ne 0$ for the region $F\leq\frac{1}{2}$. This contradicts
the fact that the Werner state is separable in the region
$F\leq\frac{1}{2}$\cite{benne}. It is why the density operator has
many possible ensembles for the given density operator. We have to
find the optimized ensemble to apply the criterion. If there
exists a pure state ensemble to make $M'$ zero, the density
operator is separable. Otherwise, the density operator is
entangled. However, this story seems to be simple in principle but it is very
complicated to find the proper ensemble practically.

We have here presented the classification and quantification scheme of a general
multipartite systems. $M'$ which is defined by the expectation values of $SU(N)$
generators determines the entangled type of multipartite states without
a priori knowledge on the quantum states investigated. However, the same entangled type
has many different states which cannot distinguish with $M'$ any more. So we have introduced
$B^{(m)}$ providing the quantification procedure. Finally, we have shown how to apply our method to the four-qubit
system as an example.

\vspace{2.0cm}

\centerline{\bf Acknowledgements}

Lee and Ahn were supported by the Korean Ministry of Science and
Technology through the Creative Research Initiatives Program under
Contact No. M10116000008-03F0000-03610. Oh was supported by
KRF-2002-070-C00029.

\end{document}